\documentclass[useAMS,a4paper,fleqn,usenatbib]{mnras}

\usepackage{newtxtext,newtxmath}

\usepackage[T1]{fontenc}
\usepackage{ae,aecompl}

\usepackage{graphicx}	% Including figure files
\usepackage{amsmath}	% Advanced maths commands
\usepackage{amssymb}	% Extra maths symbols
\usepackage{color}

\newcommand{\be}{\begin{equation}}
\newcommand{\ee}{\end{equation}}
\newcommand{\bea}{\begin{eqnarray}}
\newcommand{\eea}{\end{eqnarray}}

\def\aap{A\&A}
\def\apjl{ApJ}

\def\apjs{ApJS}
\def\apj{ApJ}
\def\araa{ARA\&A}
\def\nat{Nat}

\def\mnras{MNRAS}

\title[FSRQs detected with SKA]{Cosmological test using the high-redshift detection rate of FSRQs 
with the Square Kilometer Array}

\author[Leaf \& Melia]{Kyle Leaf$^{1}$\thanks{kyleaf@email.arizona.edu} and
Fulvio Melia$^{2}$\thanks{John Woodruff Simpson Fellow. E-mail: fmelia@email.arizona.edu} \\
$^1$Department of Physics, The University of Arizona, AZ 85721, USA \\
$^2$Department of Physics, The Applied Math Program, and Department of Astronomy, 
The University of Arizona, AZ 85721, USA}

\date{Accepted XXX. Received YYY; in original form ZZZ}

\pubyear{2019}

\begin{document}
\label{firstpage}
\pagerange{\pageref{firstpage}--\pageref{lastpage}}
\maketitle

% Abstract of the paper
\begin{abstract}
We present a phenomenological method for predicting the number of Flat Spectrum Radio Quasars 
(FSRQs) that should be detected by upcoming Square Kilometer Array (SKA) SKA1-MID Wide Band 1 and 
Medium-Deep band 2 surveys. We use the Fermi Blazar Sequence and mass estimates of Fermi FSRQs, 
and $\gamma$-ray emitting Narrow Line Seyfert 1 galaxies, to model the radio emission of FSRQs 
as a function of mass alone, assuming a near-Eddington accretion rate, which is suggested by
current quasar surveys at $z\gtrsim 6$. This is used to determine the smallest visible black
hole mass as a function of redshift in two competing cosmologies we compare in this paper: the 
standard $\Lambda$CDM model and the $R_{\rm h}=ct$ universe. We then apply lockstep growth 
to the observed black-hole mass function at $z=6$ in order to devolve that population to higher 
redshifts and determine the number of FSRQs detectable by the SKA surveys as a function of 
$z$. We find that at the redshifts for which this method is most valid, $\Lambda$CDM predicts 
$\sim 30$ times more FSRQs than $R_{\rm h}=ct$ for the Wide survey, and $\sim 100$ 
times more in the Medium-Deep survey. These stark differences will allow the SKA surveys to 
strongly differentiate between these two models, possibly rejecting one in comparison with
the other at a high level of confidence.
\end{abstract}

% Select between one and six entries from the list of approved keywords.
% Don't make up new ones.

%%%%%%%%%%%%%%%%%%%%%%%%%%%%%%%%%%%%%%%%%%%%%%%%%%

%%%%%%%%%%%%%%%%% BODY OF PAPER %%%%%%%%%%%%%%%%%%

\section{Introduction}

The Square Kilometer Array (SKA) is a premier upcoming radio observatory, expected to provide 
a powerful probe of cosmological features (Square Kilometre Array Cosmology Science Working Group, 2018, hereafter SKA CSWG 2018). Phase 1 of the SKA, constituting roughly 10\% of the design collecting area, is
expected to be constructed from 2020 to 2025.  Of principal interest for the
work reported in this paper, SKA will significantly enhance our ability to detect high-redshift, 
radio-loud quasars. The primary goal of this analysis is to calculate the number of such
sources that ought to be detectable by Phase 1 of the SKA in two competing models---standard $\Lambda$CDM 
and the $R_{\rm h}=ct$ universe---thereby providing a new test of the background cosmology.

The current standard model has been quite successful in describing many evolutionary aspects of the 
Universe, though we have begun to see growing tension between its predictions and
several key observations. For example, the Hubble constant measured with {\it Planck} (Planck
Collaboration 2018) disagrees with the locally measured value at over $4\sigma$ significance
(Riess et al. 2018). A very different type of measurement, based on the angular-diameter
distance, shows that the peak of this integrated measure also disagrees with our expectation
from $\Lambda$CDM (Melia \& Yennapureddy 2018a). All in all, comparative tests between the
standard model and the alternative picture we shall consider in this paper have now been
completed using over 25 different kinds of data. The standard model has been disfavoured by
the observations in all of these cases (see, e.g., Table~2 in Melia 2018). We
are therefore motivated to at least consider modifications to the background cosmology, perhaps
eventually even replacing it, in order to determine whether different physics may better describe 
the growing observational evidence.

The SKA, with its unprecedented capabilities, is poised to probe the intermediate to high-redshift 
radio universe with greater sensitivity than any instrument before it. In our previous study of 
high-redshift active galaxies (Fatuzzo \& Melia 2017), we analyzed the high-redshift quasar detection 
rate based on their X-ray emission in order to predict the various quasar counts to be compiled
by the upcoming eROSITA and ATHENA observatories. While $R_{\rm h}=ct$ and $\Lambda$CDM are very 
similar in describing the local universe, their predictions diverge towards higher redshifts. 
Of direct relevance to this study are the differences in the two models between the comoving 
volume and age of the Universe as functions of redshift. In our previous work, the combination 
of those factors, based on the existing observations of quasars at $z\sim6$, resulted in 
predictions of $\sim 0.16$ X-ray emitting quasars detectable by ATHENA at $z\sim7$ in $R_{\rm h}=ct$, 
compared to $\sim 160$ such objects in $\Lambda$CDM. Such a stark difference will be easily
measurable by these upcoming ATHENA observations. In this paper, we extend this work significantly
by considering the radio-loud quasar population, which will allow us to probe the Universe
even more deeply with SKA. Indeed, we shall demonstrate that SKA1's medium-deep survey has the 
potential of detecting jetted AGN with central black-hole masses as small as $M\sim10^{6}{M_\odot }$ 
out to a redshift of $z=9$. 

But our chief concern here is not only to estimate how many new radio-loud quasars we can hope
to see with SKA in the remote Universe; we shall argue that their detection rate will be strongly
dependent on the assumed background cosmology---offering us the interesting prospect of using
these counts as a powerful test of the cosmology based on techniques that are quite different from
the more conventional, integrated measures, such as luminosity distances and the Hubble parameter.
We shall seek to characterize the radio emission profile of a specific class of high-redshift 
active galaxies and evolve the population known at $z\sim6$ to higher redshifts, assuming standard 
Eddington-limited accretion, in order to predict their detection rate as a function of $z$ 
for each cosmological model. 

In \S~II, we shall discuss the radio emission characteristics of 
these high-redshift quasars, and describe the methods used to estimate their growth and evolution 
with redshift in each of these models in \S~III. To do this, we shall also summarize here the
key differences between $R_{\rm h}=ct$ and $\Lambda$CDM to the extent that these variations 
affect the AGN population synthesis. We shall present and discuss our results in \S~IV, and
finally conclude in \S~V.

\section{The radio emission of high-$z$ quasars}
As of today, approximately half of the high-$z$ (i.e., $z>5.5$), radio-loud quasars have been 
identified as Flat Spectrum Radio Quasars (FSRQs), with the remaining half showing steeper spectra 
similar to Faranov-Riley II (FR-II) galaxies (Coppejans et al. 2016). FSRQs are generally considered 
to be active galactic nuclei (AGN) with powerful relativistic jets in a configuration such that the 
jet is aligned towards the observer, commonly referred to as a blazar (Beckmann \& Shrader 2012). 
Blazars also include BL Lacertae objects (see, for example, the review papers by Antonucci 1993, and 
Sikora \& Madejski 2001). These, however, have not 
been detected at high redshifts. As discussed in Ghisellini et al. (2009a), another 
distinguishing feature between BL Lacs and FSRQs is that the former have significantly lower accretion 
rates while, in the high-redshift ($z>5.5$) Universe, the central black holes of FSRQs are found to be 
emitting at or near the Eddington luminosity (Willott et al. 2010). Circumstantial evidence, largely based
on superluminal motion (Hovatta et al. 2009), suggests that the relativistic jets of FSRQs have bulk 
Lorentz factors $\Gamma\sim5-15$ (Ghisellini \& Sbarrato, 2016). 

Assuming a random distribution of AGN orientation, the probability of seeing any given jetted (and thus, 
radio-loud) quasar in its blazar configuration is  $\sim 1/2\Gamma^2$, implying roughly a range of
$1$ in $50$ to $1$ in $450$ (Volonteri et al. 2011).  While this low blazar fraction is consistent with 
observations of AGNs at low $z$ (i.e., $z<3$), it stands in sharp contrast to the roughly $1$ in $2$ chance 
of a radio-loud, $z>4.5$ AGN being an FSRQ (Coppejans et al. 2016). Several mechanisms have been proposed 
to help explain this observation (Fabian et al. 2014; Ghisselini \& Sbarrato 2016; Wu et al. 2017), but 
analyses have been inconclusive. These proposals generally suggest that the fraction of jetted AGNs at 
high redshift that appear as FSRQs are the same as their lower-redshift counterparts, but that those 
without the jet preferentially directed at the observer are not observed strongly in the radio band.

The total spectral energy distribution (SED) across all wavebands for an AGN depends on a large number of 
physical phenomena. The more commonly included components of quasar emission are: a relativistic jet
emitting in the radio (Urry \& Padovani, 1995) and, sometimes, in X-rays and $\gamma$-rays via
inverse Compton scattering (Melia \& K\"onigl 1989); a thermally-radiating
accretion disk emitting predominantly at UV wavelengths (Shakura \& Sunyaev 1973); a hot corona emitting 
primarily in X-rays (Lusso \& Risalti 2016; Fatuzzo \& Melia 2017), and an obscuring 
torus (Krolik \& Begelman 1986); as well as any radiation from the stars in the rest of the galaxy. 
Roughly $8.1_{-3.2}^{+5.0}\%$ (Ba\~nados et al. 2015) of 
the AGNs at high redshift are known to emit strongly at radio frequencies.
 Disentangling all of the various contributions to the overall SED can be quite
challenging, especially given the diversity of AGNs sub-samples distinguished primarily upon the dominant
contributions from these many components. Rather than constructing a model based on this comprehensive
emission picture, we therefore take a more phenomenological approach, analyzing one specific class of quasar that 
ought to be the most relevant for their detection by the upcoming SKA surveys.

Our empirical approach is geared towards modeling the appearance of high-$z$ FSRQs in order to estimate 
the number of such objects that will be visible to SKA. In spite of the different population statistics 
referenced previously, FSRQs at high redshift appear to have very similar SEDs to those at lower redshift. 
For instance, the confirmed highest-redshift FSRQ to date, J0906+6930 at $z=5.47$, has an observed SED 
similar to lower-$z$ FSRQs (An \& Romani 2018). It is therefore reasonable to suppose that the large 
suite of lower-redshift FSRQ observations may be used to characterize the radio emission profile of
such objects even at higher redshifts.

Ghisellini et al. (2017) presented the so-called blazar sequence based on the Fermi observations 
compiled in Ackermann et al. (2015), an attempt to characterize the average SED of blazars separated into $\gamma$-ray 
luminosity bins. That work included separate analyses of BL Lac objects and FSRQs. As no BL Lacs have been 
observed at high redshifts, we here limit ourselves to the FSRQ data, in which the blazars are grouped into 
five bins based on their $\gamma$-ray luminosity, $L_\gamma$. These authors included the observation that
the overall SED of FSRQs does not change significantly with changing $L_\gamma$, instead generally 
differing only in terms of emitted power. Furthermore, all FSRQs in each bin were found to be consistent 
with a radio spectral index $\alpha_R\approx -0.1$, a parameter that was then held fixed in the fitting.
The results presented in Ghisellini et al. (2017) are based on an assumed $0.7=h=\Omega_\Lambda$ 
cosmology. Therefore, in order to extract the results relevant for $R_{\rm h}=ct$, we recalibrate the 
inferred luminosity using the ratio $(d_L^{R_{\rm h}=ct}/d_L^{\Lambda{\rm CDM}})^2$ 
to account for the differences between these models, in terms of the luminosity distance
$d_L$. For objects in the 3LAC catalog, the inferred luminosities are $\sim$15-19\% smaller 
in $R_{\rm h}=ct$ than in the fiducial $\Lambda$CDM model.

\begin{table*}
 \small
  \caption{Fermi blazar sequence bins, their $\gamma$-ray luminosities, and their fitted masses.}
  \centering
  \begin{tabular}{lcccc}
&& \\
    \hline
\hline
&& \\
$\log_{10} L_\gamma$ & $\log_{10} L_\gamma^{\Lambda\rm CDM}$ & $\log_{10} M^{\Lambda\rm CDM}$ & 
$\log_{10} L_\gamma^{R_{\rm h}=ct}$ & $\log_{10} M^{R_{\rm h}=ct}$\\
(${\rm erg\;s}^{-1}$) & (${\rm erg\;s}^{-1}$) & ($M_\odot$)& (${\rm erg\;s}^{-1}$)& ($M_\odot$)\\
&& \\
\hline
&& \\
$48<L$ & 48.27&$ 9.28 \pm 0.10$ & 48.19 &$ 9.24 \pm 0.10 $\\
$47<L<48$& 47.43& $8.84 \pm 0.07$ & 47.34 & $8.80 \pm 0.07$ \\
$46<L<47$ &  46.54& $8.39 \pm 0.07$ & 46.64& $ 8.34 \pm 0.07$ \\
$45<L<46$ &45.59& $7.90 \pm 0.10 $& 45.59 & $7.87 \pm 0.10$ \\
$44<L<45$& 44.59& $7.39 \pm 0.14 $& 44.59 & $7.36 \pm 0.14$ \\
&& \\
\hline\hline
  \end{tabular}
\end{table*}

For each of the five phenomenological SEDS from Ghisellini et al. (2017), we determine the 
$\gamma$-ray luminosity over the Fermi band (0.1 GeV--100 GeV) for each, and compare these results to the 
catalog of bright Fermi blazars in Ghisellini et al. (2010). This sample includes 22 FSRQs with both observed 
$\gamma$-ray luminosities and published central black-hole mass estimates. The masses are determined 
by associating an optical-UV bump in each object's SED with direct emission from the accretion disk. Ghisellini 
et al. (2009b) and Ghisellini et al. (2010) suggest that this method is accurate to within a factor of 2.
The $\nu F_\nu$ peak of the disk luminosity is used to determine the maximum temperature of the accretion disk,
which scales as $T_{\rm max} \propto (\lambda_{\rm Edd})^{1/4}M^{-1/4}$, where $\lambda_{\rm Edd}$ is the Eddington 
factor (i.e., the ratio of luminosity to its Eddington value). Furthermore, the total observed flux from the 
accretion disk scales as $\lambda_{\rm Edd}M/{d_L^2}$. Therefore, each mass determined 
by Ghisellini et al. (2010) must be multiplied by the factor $d_L^{R_{\rm h}=ct}/ 
d_L^{\Lambda{\rm CDM}}$ in order to get the appropriate inferred mass for the $R_{\rm h}=ct$ universe.
The inferred Eddington ratio in $R_{\rm h}=ct$ changes by the same factor.

These masses recorded in Ghisellini et al. (2010) (quoted as $\log_{10}[M/{M_\odot}]$) 
range from 8.17 to 9.78, with luminosities ($\log_{10}[{L_\gamma}/{\rm{erg\;}}{{\rm{s}}^{-1}}]$) from 45.93 
to 49.10, at redshifts from $z=0.213$ to $2.19$. Using the fitted SEDs from Ghisellini et al. (2010), we 
estimate the average disk luminosity of these observed FSRQs as a ratio of the Eddington limit, yielding a 
value $\approx 0.22$, providing some
evidence that the accretion rates are generally high. Since the lowest $\gamma$-ray bin in the Fermi blazar 
sequence of FSRQs falls in the range $10^{44}-10^{45}$ erg s$^{-1}$, below the range of the bright Fermi
blazars, we need an additional sample of high accretion rate active galaxies in order to associate that bin
with a central black-hole mass.

With high redshift FSRQs known to have generally high accretion rates, in order to associate the 
lower $\gamma$-ray luminosity bins of the blazar sequence with some fiducial central black-hole mass, 
we must select those objects among this category that themselves have a high accretion rate, but also 
lower luminosity. Gamma-ray emitting Narrow Line Seyfert 1 galaxies ($\gamma$-NLSy1) are radio-loud 
active galaxies characterized by flat radio spectra, exhibiting a core-jet structure, high brightness 
temperature, and apparent superluminal motion. Indeed, the broad-band SEDs of $\gamma$-NLSy1 resemble 
those of FSRQs, but merely less powerful (Paliya et al. 2013). More recent analyses of these objects 
(Paliya et al. 2018; Paliya et al. 2019) have further established a link between $\gamma$-NLSy1s and 
FSRQs. For example, attempts to fit their SEDs have shown that their jets have bulk Lorentz factors 
in the range $7<\Gamma<17$, comfortably within the usual estimated factors of FSRQs, as noted above.

\begin{figure}

	\includegraphics[width=\columnwidth]{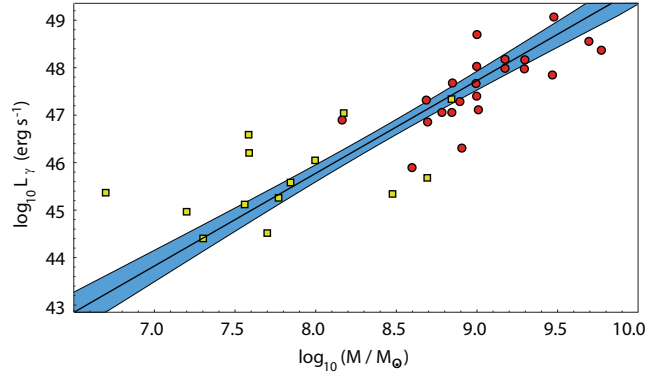}
    \caption{$\gamma$-ray luminosity $\log_{10}(L/[{\rm erg\;s}^{-1}])$ vs. central black-hole mass $\log_{10}(M/M_\odot)$ 
       in $\Lambda$CDM for bright Fermi FSRQs (red circles) (Ghisellini et al. 2009), and the 14 $\gamma$-NLSy1s 
       (yellow squares) with measured $\gamma$-ray luminosity and central black-hole mass in the 
       context of $\Lambda$CDM. (Paliya et al. 2018, 2019). The line shows the best-fit power law, along with
       its estimated uncertainty (blue swath) based on standard error propagation.}
\end{figure}

For our phenomenological model construction, we therefore use the 14 known $\gamma$-NLSy1's included 
in Paliya et al.'s (2018) sample. We take their reported $\gamma$-ray luminosity and the mass estimates 
derived from optical spectroscopy recorded in Paliya et al. (2019), and citations recorded therein. 
In $R_{\rm h}=ct$, these mass estimates are adjusted in the same manner as 
FSRQs. This sample includes reported luminosities $\log_{10}(L_\gamma/[{\rm erg\;s}^{-1}])$ 
from 44.4 to 47.4, and estimated black-hole masses from ${10^{6.7}} - {10^{8.85}}{M_ \odot }$. Together 
with the highly luminous Fermi FSRQs from Ghisellini et al. (2010), we therefore have a sample of 36 active galaxies 
with observed gamma-ray luminosities, mass estimates, and a generally high estimated inferred accretion 
rate, consistent with those $z>5.5$ quasars observed to date. Figure 1 shows the $\gamma$-ray luminosity 
plotted against the central black-hole mass for this sample, as well as our best-fit (power-law) function 
to describe their correlation. 

This best-fit power law is found with linear regression using the logrithms of the
estimated central black-hole mass and luminosity, which is taken as the independent variable.
The error in the mean mass is estimated according to 
\begin{equation}
\delta \mu \left( {{l_\gamma }} \right) = \frac{{{\sigma _\mu }}}{{\sqrt n }} \times 
\sqrt {1 + \frac{l_\gamma - \langle l_\gamma\rangle}{\sigma^2_{l_\gamma}}}\;, 
\end{equation}
where $\mu\equiv\log_{10}m$, $l_\gamma\equiv\log_{10}(L_\gamma/[{\rm erg\;s}^{-1}])$, 
$n$ is the number of observations, $\sigma_\mu$ is the square root of the reduced $\chi^2/{\rm dof}$, 
$\sigma^2_{l_\gamma }$ is the population variance of $l_\gamma$, and $\langle l_\gamma\rangle$ is the 
mean of all $l_\gamma$ values. Table 1  shows the five SED bins from Ghisellini et al. (2017), 
the $\gamma$-ray luminosity found by integrating the SED from 0.1 GeV to 100 GeV (we point out that 
integrating from 0.1 GeV to 300 GeV instead does not change the numerical result by more than $3.7\%$ or 
0.016 dex), and the mass we associate with each luminosity in both models. The errors in 
the masses are simply those described by Equation~(1) for the given $L_\gamma$.

The Fermi Blazar Sequence of Ghisellini et al. (2017) models their complete non-thermal SED in terms of 
two broad humps and a flat radio spectrum. Beginning at the lowest energies, the sequence starts with 
the flat radio segment from arbitrarily low frequencies up to some cutoff $\nu_t$. Above this, the single power
law connects with two smoothly broken power laws. The lower frequency power law is associated with
the synchrotron flux, and the higher frequency is associated with inverse Compton flux. Altogether, there
are 11 parameters determined by observations. The first set contains three typical frequencies: $\nu_t$ is where
the flat radio spectrum ends; $\nu_S$ is the peak of the synchrotron spectrum; and $\nu_C$ is the peak 
of the high-energy spectrum. Second, there are two cut-off frequencies at which the power laws start to
exhibit exponential decay. These are $\nu_{\rm{Cut},S}$, and $\nu_{\rm{Cut},C}$. Third are the four power-law
indices: $\alpha_R$ for the radio band; $\alpha_1$ connecting $\nu_t$ and $\nu_S$; $\alpha_2$ giving the
downward slope after both the $\nu_S$ and $\nu_C$ peaks, and $\alpha_3$ yielding the upward slope toward the
$\nu_C$ peak. The remaining two parameters are the luminosities at the two peaks, which are 
included as $\nu_SL(\nu_S)$ (the luminosity at the Synchrotron peak) and ratio of the 
$\nu L_\nu$ Compton and synchrotron luminosities, known as the Compton Dominance (CD). 

Having associated each Fermi blazar sequence SED with a central black hole of some mass undergoing 
Eddington-limited accretion, we interpolate the parameters characterizing the five models, resulting 
in a phenomenological SED for FSRQs as a function of $M$.
Following Ghisellini et al. (2017), we hold the radio spectral index constant at $\alpha_R=-0.1$ throughout this work. 
 For this interpolation, we additionally hold 
${\nu_{{\rm{cut,\; S}}}}$ and ${\nu_{{\rm{cut,\; C}}}}$ constant at $10^{16}\rm{\; Hz}$ and 
$10^{27}\rm{\; Hz}$, respectively, as they were inferred to be relatively constant across the sequence
models and do not significantly affect the overall SED. The indices $\alpha_1$ and $\alpha_2$ show no 
significant evolution with luminosity bin, so we hold them fixed at 0.5 and 1.43, respectively, 
while $\alpha_3$ is fitted by a linear
relation in $\log M$. The frequencies $\nu_t$ and $\nu_S$ are fitted by linear functions in $\log M$ for 
the lower bins, and both are held at $10^{12}$ Hz for the brightest two bins. The frequency $\nu_C$ is fitted with 
a quadratic interpolation between the lowest three bins, and held fixed for $M>10^{8.23}{M_\odot }$,
while $\nu_{SL}$ and CD are interpolated with quartic polynomial functions of $\log M$.

The final interpolated blazar sequence may be used to estimate the emission profile of any FSRQ 
within any waveband, from radio to $\gamma$-rays. We use the mass-dependent SED 
function to find a simple relationship between an FSRQ's
central black-hole mass and its radio luminosity, which we then extrapolate downward to smaller masses.
To do this, we switch from the full interpolated SED to a simplified relationship 
covering only the radio emission below ${10^{7.39}}{M_ \odot }$ in $\Lambda$CDM, and ${10^{7.36}}{M_ \odot }$ in 
$R_{\rm h}=ct$. We fit a power law to the three lower-luminosity 
SEDs, and find a relationship ${L_R} \propto {M^{1.55}}$ in $\Lambda$CDM, and ${L_R} \propto {M^{1.56}}$ in 
$R_{\rm h}=ct$. Even at a redshift of 7, an FSRQ matching 
the SED in the lowest bin of the blazar sequence would have a radio flux density in the SKA1 band of
$300\;\mu{\rm{Jy}}$ in $\Lambda$CDM, and $269\;\mu{\rm{Jy}}$ in $R_{\rm h}=ct$, significantly above 
the $22.8\;\mu{\rm{Jy}}$ flux limit of the Wide Band 1 survey (SKA CSWG 2018), requiring some 
extrapolation to lower masses in order to estimate the detection rate of 
smaller---though more numerous---AGNs.

\section{Calculations}
Now that the blazar radio emission may be estimated as a function of black-hole mass, we use the 
distribution from Willott et al. (2010) to establish (i.e., normalize) the mass
function at a known redshift. Following Fatuzzo \& Melia (2017), we begin with the observed 
black-hole mass function at $z=6$, and devolve the entire population 
in lockstep towards higher redshifts, with the assumption that all quasars at $z=6$ are undergoing
Eddington-limited accretion, with a duty cycle very close to 1. This growth with a fiducial efficiency 
of $10\%$ corresponds to an $e$-folding time of $t_{\rm Edd}\approx 45$ Myr (known as the Salpeter time;
see, e.g., Melia 2013). The overall growth rate of quasars at high redshifts is therefore highly 
dependent on the cosmology, given that the predicted timeline $t(z)$ changes considerably between
models.

The two competing cosmological models we consider here are the standard flat $\Lambda$CDM model 
and the $R_{\rm h}=ct$ universe. As noted earlier, while $\Lambda$CDM continues to account reasonably 
well for a broad range of data, several significant tensions continue to grow. For instance, the 
final results of the Planck mission (Planck Collaboration 2018) suggest a Hubble Constant 
statistically incompatible (now over $4\sigma$) with that inferred by Hubble Space Telescope data 
(Riess et al. 2018). In addition---and a primary motivator for this work---Weigel at al. (2015) have
estimated that, in the context of $\Lambda$CDM, $\sim 20$ AGNs should have been observed at $z>5$ in the 
Chandra Deep Field South survey, while none were identified---strongly favouring $R_{\rm h}=ct$
over the standard model (Fatuzzo \& Melia 2017). Though possible systematic effects
may be at least partially responsible for this, at face value we estimate that this negative result 
constitutes a roughly $\sim 20/\sqrt{20}$ effect, i.e., another $\sim 4\sigma$ disparity between the 
predictions of the standard model and actual observations. And more recently, Wang et al. (2018) 
 derived the quasar luminosity function at $z\sim6.7$, which
shows a significantly steeper reduction of the comoving quasar number density as a function of 
redshift than earlier estimates by Fan et al. (2001) based on $\Lambda$CDM. Of utmost importance 
in calculating this change in quasar number density as a function of redshift is the amount of 
time elapsed from one redshift to another, which is strongly dependent on the presumed cosmological
model.

For example, $z=6$ corresponds to $\approx 0.92$ Gyr after the Big Bang in flat $\Lambda$CDM with 
parameters $\Omega_{\rm m}=0.3$, ${H_0} = 70$ km s$^{-1}$ Mpc$^{-1}$. We adopt this parametrization
for the entire analysis. By comparison, it is $\approx 2$ Gyr 
in $R_{\rm h}=ct$ with the same Hubble constant (which is the only free parameter in this model). 
The $R_{\rm h}=ct$ universe is a Friedman-Lema\^itre-Robertson-Walker (FLRW) cosmology with zero active 
mass, $\rho+3p=0$ (Melia 2003, 2007, 2016, 2017, 2018; Melia \& Abdelqader 2009; Melia \& Shevchuk 2012). 
As noted in the introduction, it has accounted for the observations better than $\Lambda$CDM in comparative 
tests, including such notable cases as cosmic chronometers (Melia \& Maier 2013; Leaf \& Melia 2017; 
Melia \& Yennapureddy 2018b); strongly-lensed galaxies (Leaf \& Melia 2018; Yennapureddy \& Melia 2018); and measurements of
the maximum angular-diameter distance (Melia 2018; Melia \& Yennapureddy 2018a). The different 
timeline in $R_{\rm h}=ct$ makes it especially attractive when considering the growth rates of 
black-hole seeds into supermassive black holes, a topic addressed in detail in Melia (2013) and
Melia \& McClintock (2015). In $\Lambda$CDM, the time elapsed between $z=7$ and $z=6$ is 167 Myr, 
compared with 249 Myr in $R_{\rm h}=ct$. With a $\sim 45$ Myr e-folding time associated with 
Eddington-limited growth based on $10\%$ efficiency (see, e.g., Fatuzzo \& Melia 2017), 
$R_{\rm h}=ct$ black holes could grow by a factor of $\sim 150$ in that redshift range, 
while they would only have had time to grow by a factor of $\sim 28$ in $\Lambda$CDM.
This choice, that all of the blazars of interest are experiencing Eddington-Limited growth is motivated by the
inferred Eddington ratio of high-z quasars (such as those in the Wilott et al. 2010 sample). Thus,
with the greater elapsed time, black holes at $z>6$ are naturally expected to have been
significantly smaller in $R_{\rm h}=ct$ (for a given redshift) than in the concordance 
$\Lambda$CDM model.

The extrapolation to lower masses brings its own issues with the Willott et al (2010) black-hole
mass function, however. This distribution is well constrained only for masses in the range 
${10^8}{M_ \odot } \lesssim M \lesssim 3 \times {10^9}{M_ \odot }$. As we shall demonstrate
below, the Willott et al. mass function, when extrapolated to lower masses, may significantly 
overestimate the number of active galaxies. We begin with the black-hole 
mass function observed at $z=6$, then devolve objects to higher redshift by assuming all central 
black holes were accreting at the Eddington rate with an efficiency of $0.1$.
Analytically, a black hole accreting with efficiency $\epsilon_r$, and with a bolometric luminosity of
some fraction of the Eddington luminosity ${\lambda _{{\rm{Edd}}}} \equiv {L_{{\rm{Bol}}}}/{L_{{\rm{Edd}}}}$
will have a mass-growth rate of
\begin{equation}
{\dot M} = \frac{{\left( {1 - {\varepsilon _r}} \right)}}{{{\varepsilon _r}}}\frac{{{\lambda _{{\rm{Edd}}}}{L_{{\rm{Edd}}}}}}{{{c^2}}} = \frac{{\left( {1 - {\varepsilon _r}} \right)}}{{{\varepsilon _r}}}\frac{{{\lambda _{{\rm{Edd}}}}}}{{{t_{{\rm{Edd}}}}}}{M}.
\end{equation}
Integrating with respect to time gives the mass of the black hole as a function of time,
\begin{equation}
{M}\left( t_z \right) = {M_0}{e^{\frac{{\left( {1 - {\varepsilon _r}} \right)}}{{{\varepsilon _r}}}\frac{{{\lambda _{{\rm{Edd}}}}}}{{{t_{{\rm{Edd}}}}}}\left( {t_z - {t_{\rm obs}}} \right)}}\;,
\end{equation}
where $t_{\rm obs}$ is the age of the universe at the observed redshift, and $t_z$ is the age of the universe at redshift $z$.

\begin{figure}
	\includegraphics[width=\columnwidth]{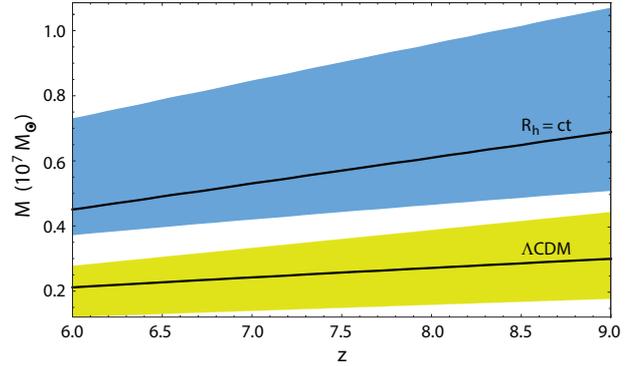}
    \caption{Smallest visible blazar in the SKA1 Wide Band 1 survey, along with our estimated
    errors: blue for $R_{\rm h}=ct$, and yellow for $\Lambda$CDM (with $\Omega_{\rm m}=0.3$). 
    The difference is due to the different radio luminosities associated with each 
    mass, and the difference in luminosity distance for each model.  The radio luminosity from 
    the Fermi blazar sequence is extrapolated to lower masses with a power law.}
\end{figure}

The black-hole mass function may then be written,
\begin{equation}
\Phi_M\left(M,z\right) = \Phi_{z=6}\left[M e^{\frac{(1-\varepsilon_r)}{\varepsilon _r}
\frac{\lambda_{\rm Edd}}{t_{\rm Edd}}(t_6 - t_z)}\right]\;,
\end{equation}
where $t_6$ and $t_z$ are the ages of the Universe, respectively, at redshifts $6$ and $z$, and $1-\epsilon_r$ gives
the fraction of infalling material accreted onto the central black hole. We 
follow Willott et al. (2010) by setting a fiducial ${\lambda_{{\rm{Edd}}}} = 1$, $\epsilon_r=0.1$  
and adopt ${t_{{\rm{Edd}}}} = 0.045$ Gyr, corresponding to the Salpeter ($e$-folding) time of 45 Myr.

The normalization $\Phi_{z=6}$ in this expression is the fitted Schechter mass function 
\begin{equation}
\Phi_{z = 6}(M) = \Phi_0\left(\frac{M}{M^*}\right)^\beta e^{-M/M^*}
\end{equation}
from Willott et al. (2010), with the best-fit parameters 
${\Phi_0} = 1.23 \times {10^{-8}}{\;\rm{Mp}}{{\rm{c}}^{-3}}{\;\rm{de}}{{\rm{x}}^{-1}}$, $M^* = 2.24 
\times {10^9}\;{M_\odot}$, and $\beta = -1.03$. This mass function is best constrained for the $z=6$ 
quasars with $M > 10^8\;M_\odot$. Note, however, that no formal error estimation was made 
in that paper. We do not attempt to estimate that error here either, but infer that it would not be 
significant to our final results compared with the error in the high-$z$ radio-loud fraction and the 
errors in associating an FSRQ central black-hole mass with a SED.

\begin{figure}
	\includegraphics[width=\columnwidth]{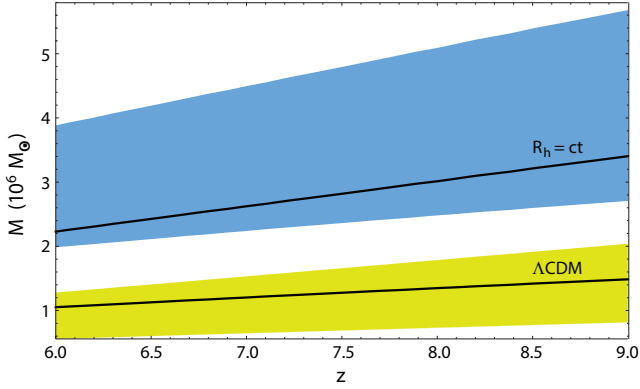}
    \caption{Same as figure 2, except for the SKA1 Medium-Deep Band 2 survey.}
\end{figure}

Over the frequency range $(0.35,1.05)$ GHz, the SKA1-Wide survey will probe 20,000 square degrees of 
sky (representing $\sim48\%$ coverage) to a sensitivity of $22.8\;\mu$Jy (SKA CSWG 2018). A $10^9\;M_\odot$ 
FSRQ at redshift $6$ is expected to have a flux $\sim 0.1$ Jy in this range, making it easily detectable. 
Likewise, the SKA1-Medium-Deep survey will probe from $0.95$ GHz to $1.75$ GHz down to its
sensitivity limit of $8.2\;\mu$Jy over 5,000 square degrees of sky (SKA CSWG 2018). 
We can therefore determine the smallest (in terms 
of black-hole mass) AGN as a function of redshift visible to the SKA1 surveys for each cosmological model. 
The SED is shifted out of the AGN's rest frame by the required factor of $1+z$, such that the luminosity 
over SKA's observable band is 
\begin{equation}
{L_R(M)} = \int\limits_{\nu_1(z + 1)}^{\nu_2(z + 1)} L_{R,\nu}(M)\;d\nu\;.
\end{equation}
The flux density in Jy over the observable band is then
\begin{equation}
F_R(M,z) = \frac{L_R(M)}{(\nu_2-\nu_1)\,4\pi d_L^2(z)}\;,
\end{equation}
where $d_L(z)$ is given in Equations~(7) and (8) as a function of $z$, and $\nu_1$ and $\nu_2$ are the
upper and lower frequencies of the observable band. In $\Lambda$CDM, we have
\begin{equation}
d_L^{\Lambda{\rm CDM}} = \frac{c}{H_0}(1+z)\int\limits_0^z \frac{du}{\sqrt{\Omega_{\rm m}(1+u)^3+
\Omega_{\rm r}(1+u)^4+\Omega_\Lambda}}\;,
\end{equation}
while in $R_{\rm h}=ct$ it takes the simpler form
\begin{equation}
d_L^{R_{\rm h}=ct} = \frac{c}{H_0}(1+z)\ln(1+z)\;.
\end{equation}

\begin{figure}
	\includegraphics[width=\columnwidth]{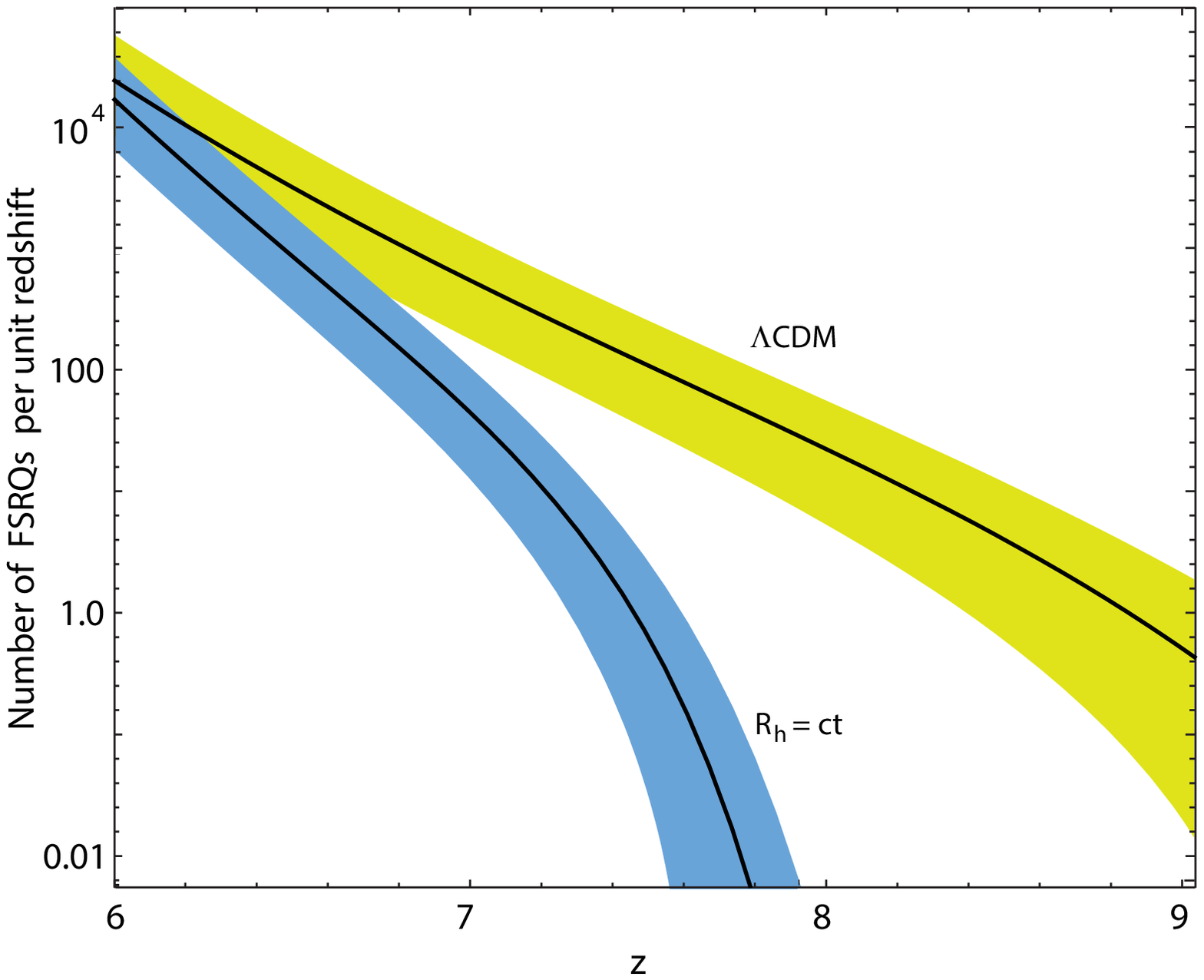}
    \caption{Estimated number of detected high-z FSRQs expected per unit redshift in
    the Wide survey, and our estimated error, for $R_{\rm h}=ct$ (blue swath) and $\Lambda$CDM 
    (yellow swath).} 
\end{figure}

\begin{figure}
	\includegraphics[width=\columnwidth]{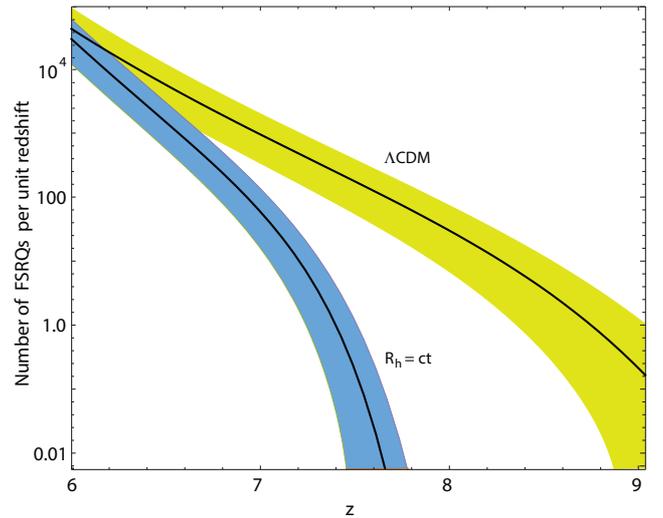}
    \caption{Estimated number of detected high-z FSRQs expected per unit redshift in the
    Medium-Deep survey, and our estimated error, for $R_{\rm h}=ct$ (blue swath) and $\Lambda$CDM 
    (yellow swath).}
\end{figure}

Next, we find the central black-hole mass that
satisfies the equality $F_{\rm lim}=F_R(M,z)$, where $F_{\rm lim}$ is the flux limit of SKA. The error in these masses is the same as the error
associated with each luminosity from Equation~(1). This error is represented by the colored swaths in figures~2 and 3.

Since the e-folding growth time is $\sim 45$ Myr, a 
$10^8\;M_\odot$ object at $z=6$ would have grown from a $5 \times 10^{6}\;M_\odot$ black hole (the smallest 
estimated central black-hole mass from our Seyfert sample) 135 Myr earlier, or from a ${2.5 \times 10^{6}}\;M_\odot$ black hole 
(the lower limit to the estimate of smallest visible quasars by the Wide survey at $z>6$) 
166 Myr earlier. In $\Lambda$CDM, these correspond to redshifts $z=6.78$ and $z=7$, respectively, 
while in $R_{\rm h}=ct$, these times are associated with $z=6.51$ and $z=6.64$. Our minimum detectable
mass estimate for the Medium-Deep survey is roughly $1.3 \times {10^6}\;M_\odot$ 
in the middle of the redshift range for $\Lambda$CDM. Thus, a ${10^8}\;M_\odot$ black
hole at $z=6$ would have passed through this minimum at redshift $z\sim7.22$ in $\Lambda$CDM and 
$z=6.76$ in $R_{\rm h}=ct$. Therefore, the predictions made in this paper are to be relied upon
most strongly for $z>6.51$  in $R_{\rm h}=ct$, and $z>6.78$ for 
$\Lambda$CDM, in the case of the Wide-field survey, and at $z>7$ and $z>7.22$, respectively, 
for the medium-deep survey.

From here, we estimate the number of objects that would be visible over the whole sky within some redshift 
bin, using the integrated number 
\begin{equation}
N = \int_{z_{\rm low}}^{z_{\rm high}}dz' \int_{\log_{10}\left[M_{\min}(z')\right]}^\infty  
\Phi(M,z')\,V_{z\,'}\,d(\log_{10} M)\;,
\end{equation}
where ${V_z} = 4\pi d_c^2\,d/dz(d_c)$ is the differential comoving volume, and ${M_{\min}}(z)$ is the 
smallest detectable black-hole mass at redshift $z$. The function plotted in figures (4) and (5) is 
\begin{equation}
dN(z) =\int_{\log_{10}\left[M_{\min}(z')\right]}^\infty  
\Phi(M,z')\,V_{z\,'}\,d(\log_{10} M)\;.
\end{equation}

The comoving distance, $d_c=d_L/(1+z)$, and 
volume are used based on the assumption that the number, $N$, of active galaxies remains constant in 
comoving volume regardless of redshift. Finally, we multiply this number by 0.0405, based on the
inferred fraction of flat-spectrum radio-loud quasars at $z>5.5$ (Ba\~nados et. al. 2015; Coppejans et al. 2016), 
and then multiply it again by the fraction of sky covered by the survey ($0.485$ for the Wide survey 
and $0.121$ for the Medium-Deep survey), to give a final estimate of the number of radio-loud 
objects expected to be visible with SKA. The results are presented in figures 4 \& 5 and Tables 2 and 3.

\begin{table*}
 \small
  \caption{Estimated counts of objects to be detected by the SKA1 Wide Band 1 survey. 
Note the anomalously high expected detections for $z<6.5$. This is due to the BH Mass 
function at $z=6$ being extrapolated to lower masses. The two rightmost columns present 
the estimates for $z$ above the redshifts for BHs in $\Lambda$CDM to have devolved such 
that a  BH at $z=6$ is smaller than the smallest known gamma-ray Radio-loud Seyfert 1 
by $z=6.78$, and below the inferred lower limit for detectable BHs at $z=7$, 
respectively.}
  \centering
  \begin{tabular}{lcccccccc}
&& \\
    \hline
\hline
&& \\
Redshift Range ($z$): & $6-6.5$ & $6.5-7$ & $7-7.5$ & $7.5-8$ & $8-8.5$ & $8.5-9$ & $6.78-9$ &$ 7-9$\\
&&&& \\
\hline
&& \\
$\Lambda$CDM & $9881_{-6193}^{+10117}$ & $1492_{-945}^{+1545}$ &  $271_{-178}^{+296}$& $51_{-36}^{+64}$ & $8.0_{-6.5}^{+13.5}$ & $0.80_{-0.75}^{+2.33} $ & $694_{-454}^{+757}$ & $330_{-221}^{+375}$\\
\\
$R_h=ct$& $5178_{-3089}^{+4785}$ & $273_{-172}^{+273}$ &  $8.1_{-6.3}^{+12.4}$& $0.01_{-0.01}^{+0.08}$ &0& 0 & $45_{-31}^{+54}$ & $8.1_{-6.3}^{+12.4}$\\
&& \\
\hline\hline
  \end{tabular}
\end{table*}

\begin{table*}
 \small
  \caption{Estimated counts of objects to be detected by the SKA1 Medium-Deep Band 2 survey. Redshift bins are the same as the Table 2, except the rightmost bin covers $z>7.22$.}
  \centering
  \begin{tabular}{lcccccccc}
&& \\
    \hline
\hline
&& \\
Redshift Range ($z$): & $6-6.5$ & $6.5-7$ & $7-7.5$ & $7.5-8$ & $8-8.5$ & $8.5-9$ & $6.78-9$ &$ 7.22-9$\\
&&&& \\
\hline
&& \\
$\Lambda$CDM & $5107_{-3341}^{+5768}$ & $786_{-517}^{+891}$ &  $151_{-101}^{+176}$& $32_{-23}^{+41}$ & $6.6_{-5.2}^{+10.3}$ & $1.1_{-1}^{+2.5} $ & $385_{-259}^{+452}$ & $96_{-67}^{+120}$\\
\\
$R_h=ct$& $2706_{-1688}^{+2741}$ & $157_{-101}^{+167}$ &  $7.2_{-5.4}^{+10.3}$& $0.05_{-0.05}^{+0.23}$ &0& 0 & $31_{-22}^{+38}$ & $1.2_{-1.1}^{+2.5}$\\
&& \\
\hline\hline
  \end{tabular}
\end{table*}
Included in each of these figures and tables are our estimated errors.
In both cases, the Hubble constant acts as an inverse length scale that defines the comoving volume,
which goes as $H_0^{-3}$. The primary source of uncertainty in our phenomenological
model lies in the error associated with the smallest detectable black-hole mass. For example, when the
best-estimate for a minimum mass is $10^6{M_\odot}$, the error from Equation~(1) is $\pm 0.28$ mass dex, or up
to $90\%$. Since the Schechter mass function predicts more numerous smaller black holes, these variations can 
drastically alter the expected object counts. As such, we use the error in the central BH mass to adjust
the lower limit of integration of Equations (10) and (11) in order to determine an upper and lower limit on the expected
all-sky objects.
This error is
\begin{eqnarray}
\delta N &=&\mp \int_{z_{\rm low}}^{z_{\rm high}}dz' \int_{\log_{10}\left[M_{\min}(z')\right]\pm\delta 
\log_{10}[M_{\min}(z')]}^{\log_{10}\left[M_{\min}(z')\right]} \Phi(M,z')\times\nonumber\\
&\null&\qquad\qquad\qquad\qquad\qquad\qquad V_{z\,'}\,d(\log_{10} M)\;,
\end{eqnarray}
with the choice in signs corresponding to the upper and lower errors.
We also include the errors in
the high-$z$ radio-loud fraction (i.e., $8.1_{-3.2}^{+5.0}\%$; Ba\~nados et al. 2015) and the blazar fraction 
($0.5\pm 0.1$) of those radio loud objects, based on the 26 objects in Coppejans et al. (2016).
Added in quadrature, we find a final fractional error in the expected counts of
\begin{equation}
\sigma_N^+\equiv {\left( {\frac{\delta N}{N}} \right)^2} + {\left( {\frac{{0.1}}{{0.5}}} \right)^2} + {\left( {\frac{{0.05}}{{0.081}}} \right)^2}
\end{equation}
at the upper end, and an analogous expression, $\sigma_N^-$ (with 0.05 replaced with 0.032) at the lower end.
These errors correspond to the colored swaths visualized in figures~4 and 5.

\section{Discussion}
The results of this analysis suggest that SKA will detect a highly significant difference in the number of 
blazars as a function of redshift for the two cosmological models we have compared in this paper. We expect to 
find far more quasars in $\Lambda$CDM than in $R_{\rm h}=ct$. The first few columns in Tables 2 and 3 point to 
the deficiencies in Willott et al.'s (2010) black-hole mass function at $z=6$. These suggest that SKA would be 
able to find nearly 10,000 radio-loud quasars in the redshift range $6\lesssim z\lesssim 6.5$, 
roughly one per two square degrees in the wide survey. But the vast majority of these are due 
to an extrapolation of the $z=6$ mass function to masses far below those used to construct this distribution. 
With no real existing data to motivate the counts of smaller black holes at this redshift, these results have 
very little predictive power. Therefore, the predictions made in this paper are reliable only at 
$z\gtrsim 7$, as discussed in \S\S~III and IV. 

The differences between the two models become more extreme with increasing redshift. At $z>7.22$, 
the standard $\Lambda$CDM model predicts that SKA will detect 80 times more blazars than 
$R_{\rm h}=ct$. The latter cosmology predicts effectively zero detections in either survey 
at $z>7.5$. The current 
most-distant known quasar of any type lies at a redshift $z=7.54$. This object has an estimated 
mass of $\sim 10^8\;M_\odot$ (Ba\~nados et al. 2018). When we begin with Willot et al.'s black-hole 
mass function at $z=6$ and apply lockstep evolution, objects such as J1342+0928 appear to be
statistically impossible for both considered cosmologies. But evidently this object and other similar 
objects at $z\gtrsim 7$ exist. Therefore, it becomes immediately clear that the simple evolution of 
the $z=6$ mass function does not accurately describe the full AGN population, if either of these 
cosmological models is to be believed. We point out, however, that the $z=6$ mass function does not 
take into account black holes that have already evolved past their primary active phase. Their host 
galaxies would then no longer appear active, and would be missed (or classified as some other kind
of object if detected) by a survey. Therefore, we suggest that the $z=6$ mass function devolved to 
higher redshifts is a {\it lower limit} to the expected counts of existing AGNs. Our predicted
numbers of FSRQs to be detected by SKA1 Wide and Medium Deep surveys thus represent lower 
limits on the number of objects that would be detected in either cosmological model. 

Nonetheless,
we point out that the number of objects predicted by $\Lambda$CDM at $z>7$ in the Wide survey is 
more than a factor of $40$ greater than the prediction by $R_{\rm h}=ct$. With the lower 
end of the $z>7$ estimate being 174, significantly above the $R_{\rm h}=ct$ range, we expect the results of 
the survey to strongly prefer one model over the other. Furthermore, $R_{\rm h}=ct$ predicts fewer than 1 
detected blazar at $z>7.5$ in both surveys, while $\Lambda$CDM still predicts $\sim~60$ 
in the wide survey, and $\sim~40$ in the medium-deep survey. Likewise, in the deep survey, $\Lambda$CDM 
predicts roughly 100 blazars at $z>7.22$, while only $\sim 1$ is expected in $R_{\rm h}=ct$. We do not 
anticipate that quasars unaccounted for in our simplified evolution picture could overcome 
this large difference between the two models.

\section{Conclusion}
We have presented a method for predicting the number of Flat Spectrum Radio Quasars based on a 
phenomenological approach to estimating the radio luminosity of high-redshift sources. Using the
observed population of AGNs with known $\gamma$-ray luminosities and mass estimates at lower
redshifts, and using the Fermi Blazar Sequence, we have constructed a broad-band spectral energy 
distribution of FSRQs as a function of mass alone. We have then used this spectrum to estimate the 
smallest black-hole mass of a FSRQ detectable by SKA as a function of redshift. Finally, we have
devolved the black-hole mass function observed at $z=6$ to higher redshifts, assuming lockstep 
evolution. 

With the assumption that all the black holes above the limiting mass at a given redshift would 
be visible to a complete flux-limited survey, we have predict the number counts of FSRQs expected 
as a function of redshift for two competing models, $\Lambda$CDM and $R_{\rm h}=ct$. Taking into 
account the limitations used to construct the black-hole mass function at $z=6$, we have predicted 
the counts of objects at $z>7$ for the wide-field SKA band 1 survey, and at $z>7.22$ in the 
medium-deep SKA band 2 survey. These results offer a definitive method of discriminating between these 
two cosmological models, in which $\Lambda$CDM predicts $30$ times more blazar counts than
$R_{\rm h}=ct$ in the wide survey, and $80$ times more in the medium-deep survey. 

While there appear to be other influences at work regarding the evolution of the black-hole mass
at high redshifts, this analysis may be viewed as providing a lower limit to the overall number
of such sources expected to be seen by SKA at those redshifts. At this time,
we do not anticipate any additional corrections in $R_{\rm h}=ct$ that could reasonably
explain the detection of $\sim 200$ such sources by the Wide survey at $ z>7$ if this is
what SKA finds. But if fewer than $\sim 50$ detections are made by the time the survey
is completed, it would correspondingly be very difficult to explain this deficit in the
context of the standard model, thereby strongly favouring $R_{\rm h}=ct$. 

We also point out that the phenomenological approach used in this paper may be applied to any 
other nonthermal waveband of interest, allowing for predictions of FSRQ-type objects by different surveys
in other parts of the spectrum. In addition, the SKA surveys are poised to discover AGNs with smaller 
black-hole masses than ever observed before at $z>6$. We expect that SKA detections at 
$z\sim6$ will thereby extend the $z=6$ black-hole mass function
to significantly lower masses, thus permitting us to eventually devolve this improved
mass distribution towards higher redshifts. Our ultimate goal is to form a sufficiently
complete picture of AGN growth in the early universe in order to convincingly rule out
one or other of these two models at a very high level of confidence.

\section*{Acknowledgments}

We thank Jinyi Yang, Xiaohui Fan, and Feige Wang for helpful discussions
concerning recent measurements of high-redshift quasars.

\nocite{*}
\bibliographystyle{mnras}
\bibliography{sources}

\begin{thebibliography}{}
\makeatletter
\relax
\def\mn@urlcharsother{\let\do\@makeother \do\$\do\&\do\#\do\^\do\_\do\%\do\~}
\def\mn@doi{\begingroup\mn@urlcharsother \@ifnextchar [ {\mn@doi@}
  {\mn@doi@[]}}
\def\mn@doi@[#1]#2{\def\@tempa{#1}\ifx\@tempa\@empty \href
  {http://dx.doi.org/#2} {doi:#2}\else \href {http://dx.doi.org/#2} {#1}\fi
  \endgroup}
\def\mn@eprint#1#2{\mn@eprint@#1:#2::\@nil}
\def\mn@eprint@arXiv#1{\href {http://arxiv.org/abs/#1} {{\tt arXiv:#1}}}
\def\mn@eprint@dblp#1{\href {http://dblp.uni-trier.de/rec/bibtex/#1.xml}
  {dblp:#1}}
\def\mn@eprint@#1:#2:#3:#4\@nil{\def\@tempa {#1}\def\@tempb {#2}\def\@tempc
  {#3}\ifx \@tempc \@empty \let \@tempc \@tempb \let \@tempb \@tempa \fi \ifx
  \@tempb \@empty \def\@tempb {arXiv}\fi \@ifundefined
  {mn@eprint@\@tempb}{\@tempb:\@tempc}{\expandafter \expandafter \csname
  mn@eprint@\@tempb\endcsname \expandafter{\@tempc}}}

\bibitem[\protect\citeauthoryear{{Ackermann} et~al.,}{{Ackermann}
  et~al.}{2015}]{2015yCat..18100014A}
{Ackermann} M.,  et~al., 2015, VizieR Online Data Catalog, \href
  {https://ui.adsabs.harvard.edu/\#abs/2015yCat..18100014A} {p. J/ApJ/810/14}

\bibitem[\protect\citeauthoryear{{An} \& {Romani}}{{An} \&
  {Romani}}{2018}]{2018ApJ...856..105A}
{An} H.,  {Romani} R.~W.,  2018, \mn@doi [\apj] {10.3847/1538-4357/aab435},
  \href {https://ui.adsabs.harvard.edu/\#abs/2018ApJ...856..105A} {856, 105}

\bibitem[\protect\citeauthoryear{{Antonucci}}{{Antonucci}}{1993}]{1993ARA&A..31..473A}
{Antonucci} R.,  1993, \mn@doi [\araa] {10.1146/annurev.aa.31.090193.002353},
  \href {http://adsabs.harvard.edu/abs/1993ARA%26A..31..473A} {31, 473}

\bibitem[\protect\citeauthoryear{{Ba{\~n}ados} et~al.,}{{Ba{\~n}ados}
  et~al.}{2015}]{2015ApJ...804..118B}
{Ba{\~n}ados} E.,  et~al., 2015, \mn@doi [\apj] {10.1088/0004-637X/804/2/118},
  \href {https://ui.adsabs.harvard.edu/\#abs/2015ApJ...804..118B} {804, 118}

\bibitem[\protect\citeauthoryear{{Ba{\~n}ados} et~al.,}{{Ba{\~n}ados}
  et~al.}{2018}]{2018Natur.553..473B}
{Ba{\~n}ados} E.,  et~al., 2018, \mn@doi [\nat] {10.1038/nature25180}, \href
  {https://ui.adsabs.harvard.edu/\#abs/2018Natur.553..473B} {553, 473}

\bibitem[\protect\citeauthoryear{{Beckmann} \& {Shrader}}{{Beckmann} \&
  {Shrader}}{2012}]{2012agn..book.....B}
{Beckmann} V.,  {Shrader} C.~R.,  2012, {Active Galactic Nuclei}

\bibitem[\protect\citeauthoryear{{Bull} et~al.,}{{Bull}
  et~al.}{2018}]{2018arXiv181002680B}
{Bull} P.,  et~al., 2018, arXiv e-prints, \href
  {https://ui.adsabs.harvard.edu/\#abs/2018arXiv181002680B} {p.
  arXiv:1810.02680}

\bibitem[\protect\citeauthoryear{{Coppejans} et~al.,}{{Coppejans}
  et~al.}{2016}]{2016MNRAS.463.3260C}
{Coppejans} R.,  et~al., 2016, \mn@doi [\mnras] {10.1093/mnras/stw2236}, \href
  {https://ui.adsabs.harvard.edu/\#abs/2016MNRAS.463.3260C} {463, 3260}

\bibitem[\protect\citeauthoryear{{Fabian}, {Walker}, {Celotti}, {Ghisellini},
  {Mocz}, {Blundell}  \& {McMahon}}{{Fabian}
  et~al.}{2014}]{2014MNRAS.442L..81F}
{Fabian} A.~C.,  {Walker} S.~A.,  {Celotti} A.,  {Ghisellini} G.,  {Mocz} P.,
  {Blundell} K.~M.,   {McMahon} R.~G.,  2014, \mn@doi [\mnras]
  {10.1093/mnrasl/slu065}, \href
  {http://adsabs.harvard.edu/abs/2014MNRAS.442L..81F} {442, L81}

\bibitem[\protect\citeauthoryear{{Fan} et~al.,}{{Fan}
  et~al.}{2001}]{2001AJ....122.2833F}
{Fan} X.,  et~al., 2001, \mn@doi [\aj] {10.1086/324111}, \href
  {https://ui.adsabs.harvard.edu/\#abs/2001AJ....122.2833F} {122, 2833}

\bibitem[\protect\citeauthoryear{{Fatuzzo} \& {Melia}}{{Fatuzzo} \&
  {Melia}}{2017}]{2017ApJ...846..129F}
{Fatuzzo} M.,  {Melia} F.,  2017, \mn@doi [\apj] {10.3847/1538-4357/aa8627},
  \href {https://ui.adsabs.harvard.edu/\#abs/2017ApJ...846..129F} {846, 129}

\bibitem[\protect\citeauthoryear{{Ghisellini} \& {Sbarrato}}{{Ghisellini} \&
  {Sbarrato}}{2016}]{2016MNRAS.461L..21G}
{Ghisellini} G.,  {Sbarrato} T.,  2016, \mn@doi [\mnras]
  {10.1093/mnrasl/slw089}, \href
  {https://ui.adsabs.harvard.edu/\#abs/2016MNRAS.461L..21G} {461, L21}

\bibitem[\protect\citeauthoryear{{Ghisellini}, {Maraschi}  \&
  {Tavecchio}}{{Ghisellini} et~al.}{2009a}]{2009MNRAS.396L.105G}
{Ghisellini} G.,  {Maraschi} L.,   {Tavecchio} F.,  2009a, \mn@doi [\mnras]
  {10.1111/j.1745-3933.2009.00673.x}, \href
  {https://ui.adsabs.harvard.edu/\#abs/2009MNRAS.396L.105G} {396, L105}

\bibitem[\protect\citeauthoryear{{Ghisellini}, {Foschini}, {Volonteri},
  {Ghirlanda}, {Haardt}, {Burlon}  \& {Tavecchio}}{{Ghisellini}
  et~al.}{2009b}]{2009MNRAS.399L..24G}
{Ghisellini} G.,  {Foschini} L.,  {Volonteri} M.,  {Ghirlanda} G.,  {Haardt}
  F.,  {Burlon} D.,   {Tavecchio} F.,  2009b, \mn@doi [\mnras]
  {10.1111/j.1745-3933.2009.00716.x}, \href
  {http://adsabs.harvard.edu/abs/2009MNRAS.399L..24G} {399, L24}

\bibitem[\protect\citeauthoryear{{Ghisellini}, {Tavecchio}, {Foschini},
  {Ghirlanda}, {Maraschi}  \& {Celotti}}{{Ghisellini}
  et~al.}{2010}]{2010MNRAS.402..497G}
{Ghisellini} G.,  {Tavecchio} F.,  {Foschini} L.,  {Ghirlanda} G.,  {Maraschi}
  L.,   {Celotti} A.,  2010, \mn@doi [\mnras]
  {10.1111/j.1365-2966.2009.15898.x}, \href
  {http://adsabs.harvard.edu/abs/2010MNRAS.402..497G} {402, 497}

\bibitem[\protect\citeauthoryear{{Ghisellini}, {Righi}, {Costamante}  \&
  {Tavecchio}}{{Ghisellini} et~al.}{2017}]{2017MNRAS.469..255G}
{Ghisellini} G.,  {Righi} C.,  {Costamante} L.,   {Tavecchio} F.,  2017,
  \mn@doi [\mnras] {10.1093/mnras/stx806}, \href
  {https://ui.adsabs.harvard.edu/\#abs/2017MNRAS.469..255G} {469, 255}

\bibitem[\protect\citeauthoryear{{Hovatta}, {Valtaoja}, {Tornikoski}  \&
  {L{\"a}hteenm{\"a}ki}}{{Hovatta} et~al.}{2009}]{2009A&A...494..527H}
{Hovatta} T.,  {Valtaoja} E.,  {Tornikoski} M.,   {L{\"a}hteenm{\"a}ki} A.,
  2009, \mn@doi [\aap] {10.1051/0004-6361:200811150}, \href
  {http://adsabs.harvard.edu/abs/2009A%26A...494..527H} {494, 527}

\bibitem[\protect\citeauthoryear{{Krolik} \& {Begelman}}{{Krolik} \&
  {Begelman}}{1986}]{1986ApJ...308L..55K}
{Krolik} J.~H.,  {Begelman} M.~C.,  1986, \mn@doi [\apjl] {10.1086/184743},
  \href {http://adsabs.harvard.edu/abs/1986ApJ...308L..55K} {308, L55}

\bibitem[\protect\citeauthoryear{{Leaf} \& {Melia}}{{Leaf} \&
  {Melia}}{2017}]{2017MNRAS.470.2320L}
{Leaf} K.,  {Melia} F.,  2017, \mn@doi [\mnras] {10.1093/mnras/stx1437}, \href
  {https://ui.adsabs.harvard.edu/\#abs/2017MNRAS.470.2320L} {470, 2320}

\bibitem[\protect\citeauthoryear{{Lusso} \& {Risaliti}}{{Lusso} \&
  {Risaliti}}{2016}]{2016ApJ...819..154L}
{Lusso} E.,  {Risaliti} G.,  2016, \mn@doi [\apj]
  {10.3847/0004-637X/819/2/154}, \href
  {http://adsabs.harvard.edu/abs/2016ApJ...819..154L} {819, 154}

\bibitem[\protect\citeauthoryear{{Melia}}{{Melia}}{2003}]{2003edin.book.....M}
{Melia} F.,  2003, {The Edge of Infinity}

\bibitem[\protect\citeauthoryear{{Melia}}{{Melia}}{2007}]{2007MNRAS.382.1917M}
{Melia} F.,  2007, \mn@doi [\mnras] {10.1111/j.1365-2966.2007.12499.x}, \href
  {https://ui.adsabs.harvard.edu/\#abs/2007MNRAS.382.1917M} {382, 1917}

\bibitem[\protect\citeauthoryear{{Melia}}{{Melia}}{2013}]{2013ApJ...764...72M}
{Melia} F.,  2013, \mn@doi [\apj] {10.1088/0004-637X/764/1/72}, \href
  {http://adsabs.harvard.edu/abs/2013ApJ...764...72M} {764, 72}

\bibitem[\protect\citeauthoryear{{Melia}}{{Melia}}{2016}]{2016FrPhy..11k9801M}
{Melia} F.,  2016, \mn@doi [Frontiers of Physics] {10.1007/s11467-016-0557-6},
  \href {https://ui.adsabs.harvard.edu/\#abs/2016FrPhy..11k9801M} {11, 119801}

\bibitem[\protect\citeauthoryear{{Melia}}{{Melia}}{2017}]{2017MNRAS.464.1966M}
{Melia} F.,  2017, \mn@doi [\mnras] {10.1093/mnras/stw2493}, \href
  {https://ui.adsabs.harvard.edu/\#abs/2017MNRAS.464.1966M} {464, 1966}

\bibitem[\protect\citeauthoryear{{Melia}}{{Melia}}{2018}]{2018MNRAS.481.4855M}
{Melia} F.,  2018, \mn@doi [\mnras] {10.1093/mnras/sty2596}, \href
  {https://ui.adsabs.harvard.edu/\#abs/2018MNRAS.481.4855M} {481, 4855}

\bibitem[\protect\citeauthoryear{{Melia} \& {Abdelqader}}{{Melia} \&
  {Abdelqader}}{2009}]{2009IJMPD..18.1889M}
{Melia} F.,  {Abdelqader} M.,  2009, \mn@doi [International Journal of Modern
  Physics D] {10.1142/S0218271809015746}, \href
  {https://ui.adsabs.harvard.edu/\#abs/2009IJMPD..18.1889M} {18, 1889}

\bibitem[\protect\citeauthoryear{{Melia} \& {Koenigl}}{{Melia} \&
  {Koenigl}}{1989}]{1989ESASP.296..995M}
{Melia} F.,  {Koenigl} A.,  1989, in {Hunt} J.,  {Battrick} B.,  eds,  ESA
  Special Publication Vol. 296, Two Topics in X-Ray Astronomy, Volume 1: X Ray
  Binaries. Volume 2: AGN and the X Ray Background. pp 995--1000

\bibitem[\protect\citeauthoryear{{Melia} \& {Maier}}{{Melia} \&
  {Maier}}{2013}]{2013MNRAS.432.2669M}
{Melia} F.,  {Maier} R.~S.,  2013, \mn@doi [\mnras] {10.1093/mnras/stt596},
  \href {https://ui.adsabs.harvard.edu/\#abs/2013MNRAS.432.2669M} {432, 2669}

\bibitem[\protect\citeauthoryear{{Melia} \& {McClintock}}{{Melia} \&
  {McClintock}}{2015}]{2015RSPSA.47150449M}
{Melia} F.,  {McClintock} T.~M.,  2015, \mn@doi [Proceedings of the Royal
  Society of London Series A] {10.1098/rspa.2015.0449}, \href
  {https://ui.adsabs.harvard.edu/\#abs/2015RSPSA.47150449M} {471, 20150449}

\bibitem[\protect\citeauthoryear{{Melia} \& {Shevchuk}}{{Melia} \&
  {Shevchuk}}{2012}]{2012MNRAS.419.2579M}
{Melia} F.,  {Shevchuk} A.~S.~H.,  2012, \mn@doi [\mnras]
  {10.1111/j.1365-2966.2011.19906.x}, \href
  {https://ui.adsabs.harvard.edu/\#abs/2012MNRAS.419.2579M} {419, 2579}

\bibitem[\protect\citeauthoryear{{Melia} \& {Yennapureddy}}{{Melia} \&
  {Yennapureddy}}{2018a}]{2018MNRAS.480.2144M}
{Melia} F.,  {Yennapureddy} M.~K.,  2018a, \mn@doi [\mnras]
  {10.1093/mnras/sty1962}, \href
  {https://ui.adsabs.harvard.edu/\#abs/2018MNRAS.480.2144M} {480, 2144}

\bibitem[\protect\citeauthoryear{{Melia} \& {Yennapureddy}}{{Melia} \&
  {Yennapureddy}}{2018b}]{2018JCAP...02..034M}
{Melia} F.,  {Yennapureddy} M.~K.,  2018b, \mn@doi [Journal of Cosmology and
  Astro-Particle Physics] {10.1088/1475-7516/2018/02/034}, \href
  {https://ui.adsabs.harvard.edu/\#abs/2018JCAP...02..034M} {2018, 034}

\bibitem[\protect\citeauthoryear{{Paliya}, {Stalin}, {Shukla}  \&
  {Sahayanathan}}{{Paliya} et~al.}{2013}]{2013ApJ...768...52P}
{Paliya} V.~S.,  {Stalin} C.~S.,  {Shukla} A.,   {Sahayanathan} S.,  2013,
  \mn@doi [\apj] {10.1088/0004-637X/768/1/52}, \href
  {https://ui.adsabs.harvard.edu/\#abs/2013ApJ...768...52P} {768, 52}

\bibitem[\protect\citeauthoryear{{Paliya}, {Ajello}, {Rakshit}, {Mand al},
  {Stalin}, {Kaur}  \& {Hartmann}}{{Paliya} et~al.}{2018}]{2018ApJ...853L...2P}
{Paliya} V.~S.,  {Ajello} M.,  {Rakshit} S.,  {Mand al} A.~K.,  {Stalin} C.~S.,
   {Kaur} A.,   {Hartmann} D.,  2018, \mn@doi [\apj]
  {10.3847/2041-8213/aaa5ab}, \href
  {https://ui.adsabs.harvard.edu/\#abs/2018ApJ...853L...2P} {853, L2}

\bibitem[\protect\citeauthoryear{{Paliya}, {Parker}, {Jiang}, {Fabian},
  {Brenneman}, {Ajello}  \& {Hartmann}}{{Paliya}
  et~al.}{2019}]{2019ApJ...872..169P}
{Paliya} V.~S.,  {Parker} M.~L.,  {Jiang} J.,  {Fabian} A.~C.,  {Brenneman} L.,
   {Ajello} M.,   {Hartmann} D.,  2019, \mn@doi [\apj]
  {10.3847/1538-4357/ab01ce}, \href
  {https://ui.adsabs.harvard.edu/\#abs/2019ApJ...872..169P} {872, 169}

\bibitem[\protect\citeauthoryear{{Planck Collaboration} et~al.,}{{Planck
  Collaboration} et~al.}{2018}]{2018arXiv180706209P}
{Planck Collaboration} et~al., 2018, arXiv e-prints, \href
  {https://ui.adsabs.harvard.edu/\#abs/2018arXiv180706209P} {p.
  arXiv:1807.06209}

\bibitem[\protect\citeauthoryear{{Riess} et~al.,}{{Riess}
  et~al.}{2018}]{2018ApJ...861..126R}
{Riess} A.~G.,  et~al., 2018, \mn@doi [\apj] {10.3847/1538-4357/aac82e}, \href
  {http://adsabs.harvard.edu/abs/2018ApJ...861..126R} {861, 126}

\bibitem[\protect\citeauthoryear{{Shakura}}{{Shakura}}{1973}]{1973SvA....16..756S}
{Shakura} N.~I.,  1973, \sovast, \href
  {https://ui.adsabs.harvard.edu/\#abs/1973SvA....16..756S} {16, 756}

\bibitem[\protect\citeauthoryear{{Sikora} \& {Madejski}}{{Sikora} \&
  {Madejski}}{2001}]{2001AIPC..558..275S}
{Sikora} M.,  {Madejski} G.,  2001, in {Aharonian} F.~A.,  {V{\"o}lk} H.~J.,
  eds,  American Institute of Physics Conference Series Vol. 558, American
  Institute of Physics Conference Series. pp 275--288 (\mn@eprint {}
  {astro-ph/0101382}), \mn@doi{10.1063/1.1370797}

\bibitem[\protect\citeauthoryear{{Square Kilometre Array Cosmology Science
  Working Group} et~al.,}{{Square Kilometre Array Cosmology Science Working
  Group} et~al.}{2018}]{2018arXiv181102743S}
{Square Kilometre Array Cosmology Science Working Group} et~al., 2018, arXiv
  e-prints, \href {http://adsabs.harvard.edu/abs/2018arXiv181102743S} {}

\bibitem[\protect\citeauthoryear{{Urry} \& {Padovani}}{{Urry} \&
  {Padovani}}{1995}]{1995PASP..107..803U}
{Urry} C.~M.,  {Padovani} P.,  1995, \mn@doi [\pasp] {10.1086/133630}, \href
  {http://adsabs.harvard.edu/abs/1995PASP..107..803U} {107, 803}

\bibitem[\protect\citeauthoryear{{Volonteri}, {Haardt}, {Ghisellini}  \& {Della
  Ceca}}{{Volonteri} et~al.}{2011}]{2011MNRAS.416..216V}
{Volonteri} M.,  {Haardt} F.,  {Ghisellini} G.,   {Della Ceca} R.,  2011,
  \mn@doi [\mnras] {10.1111/j.1365-2966.2011.19024.x}, \href
  {http://adsabs.harvard.edu/abs/2011MNRAS.416..216V} {416, 216}

\bibitem[\protect\citeauthoryear{{Wang} et~al.,}{{Wang}
  et~al.}{2018}]{2018arXiv181011926W}
{Wang} F.,  et~al., 2018, arXiv e-prints, \href
  {https://ui.adsabs.harvard.edu/\#abs/2018arXiv181011926W} {p.
  arXiv:1810.11926}

\bibitem[\protect\citeauthoryear{{Weigel}, {Schawinski}, {Treister}, {Urry},
  {Koss}  \& {Trakhtenbrot}}{{Weigel} et~al.}{2015}]{2015MNRAS.448.3167W}
{Weigel} A.~K.,  {Schawinski} K.,  {Treister} E.,  {Urry} C.~M.,  {Koss} M.,
  {Trakhtenbrot} B.,  2015, \mn@doi [\mnras] {10.1093/mnras/stv184}, \href
  {https://ui.adsabs.harvard.edu/\#abs/2015MNRAS.448.3167W} {448, 3167}

\bibitem[\protect\citeauthoryear{{Willott} et~al.,}{{Willott}
  et~al.}{2010}]{2010AJ....140..546W}
{Willott} C.~J.,  et~al., 2010, \mn@doi [\aj] {10.1088/0004-6256/140/2/546},
  \href {https://ui.adsabs.harvard.edu/\#abs/2010AJ....140..546W} {140, 546}

\bibitem[\protect\citeauthoryear{{Wu}, {Ghisellini}, {Hodges-Kluck}, {Gallo},
  {Ciardi}, {Haardt}, {Sbarrato}  \& {Tavecchio}}{{Wu}
  et~al.}{2017}]{2017MNRAS.468..109W}
{Wu} J.,  {Ghisellini} G.,  {Hodges-Kluck} E.,  {Gallo} E.,  {Ciardi} B.,
  {Haardt} F.,  {Sbarrato} T.,   {Tavecchio} F.,  2017, \mn@doi [\mnras]
  {10.1093/mnras/stx416}, \href
  {https://ui.adsabs.harvard.edu/\#abs/2017MNRAS.468..109W} {468, 109}

\bibitem[\protect\citeauthoryear{{Yennapureddy} \& {Melia}}{{Yennapureddy} \&
  {Melia}}{2018}]{2018EPJC...78..258Y}
{Yennapureddy} M.~K.,  {Melia} F.,  2018, \mn@doi [European Physical Journal C]
  {10.1140/epjc/s10052-018-5746-8}, \href
  {http://adsabs.harvard.edu/abs/2018EPJC...78..258Y} {78, 258}

\makeatother
\end{thebibliography}

%%%%%%%%%%%%%%%%%%%%%%%%%%%%%%%%%%%%%%%%%%%%%%%%%%

% Don't change these lines
\bsp	% typesetting comment
\label{lastpage}
\end{document}